\documentstyle[12pt]{article}
\newcommand{\tvi}{\vrule height 12pt depth 5pt width 0pt}
\newcommand{\bcn}{\begin{center}}
\newcommand{\beq}{\begin{equation}}
\newcommand{\beqn}{\begin{eqnarray}}
\newcommand{\ecn}{\end{center}}
\newcommand{\eeq}{\end{equation}}
\newcommand{\eeqn}{\end{eqnarray}}
\def\slash{\llap{/}}
\def\lapprox{\mathrel{\mathop{\kern 0pt <}\limits_{\displaystyle\sim}}}
\def\gapprox{\mathrel{\mathop{\kern 0pt >}\limits_{\displaystyle\sim}}}
\begin{document}
\newpage
\setlength{\topmargin}{4.0cm}

\begin{center}
{\Large {\bf HARD EXCLUSIVE SCATTERING IN QCD
\footnote{\rm to be published in the proceedings of the ELFE summer school on
confinement physics, Cambridge (1995)}
}} \\
\vspace{1.5cm}
{\large Thierry GOUSSET$^{a,b}$ and Bernard PIRE$^{b}$} \\
\medskip
{\it a. CEA, Service de Physique Nucl\'eaire/DAPNIA, CE Saclay, F91191 Gif,
France\\
b. Centre de Physique Th\'eorique{\footnote {Unit\'e propre 14 du Centre
National
de la Recherche Scientifique.}}, Ecole Polytechnique, F91128 Palaiseau,
France} \\

\vspace{4.0cm}

\large {\bf ABSTRACT}

\end{center}


We review the theory of hard exclusive scattering in Quantum Chromodynamics.
 After recalling the classical counting rules which describe the leading scale
dependence of form factors and elastic reactions at fixed angle, the
 pedagogical example of the
pion form factor is developped in some detail in order to show explicitely what
factorization means in the QCD framework. The inclusion of transverse degrees
of
freedom leads to the discussion of Sudakov effects which are crucial for
protecting
the calculation from dangerous infrared regions. The picture generalizes to
many
hard reactions; a strategy to extract distribution amplitudes from future data
is sketched. We discuss also the particular case of hadron-hadron
collisions where the
independent scattering mechanism dominates asymptotically and where a different
factorization formula applies. We briefly present the concepts of color
transparency
and nuclear filtering and critically discuss the few present data on this
subject.

\vfill

\pagebreak

\setlength{\topmargin}{-1.5cm}
\setlength{\textheight}{25.0cm}
\setlength{\baselineskip}{0.75cm}

We consider here {\it exclusive processes }, that is
interactions  resulting in a  final state where all
particles are identified. Using a
perturbative expansion to study these reactions may {\it a priori}
be foreseen if a large momentum transfer appears: this is what is called a
{\it hard} reaction.
Let us take as an example  Compton scattering on a nucleon. The
process is
\beq
\gamma(k)+N(p)\rightarrow\gamma(k')+N(p'),
\eeq
and we are interested in the polarized or unpolarized differential cross
section
\beq
{d\sigma\over dt}(s,t),\hskip 5mm s=(k+p)^2,\hskip 5mm t=(k-k')^2,
\eeq
for large values of  $|t|\sim s$ (large means much bigger than
hadronic or confinement scales).
In the ultra-relativistic limit, one may  neglect the nucleon mass,
and kinematics simplifies. In the center of  mass frame, one has
\beq
E={\sqrt{s}\over 2},\ \sin^2{\theta\over 2}=-{t\over s},
\eeq
where $E$ is the energy of any particules and $\theta$ is the angle between
the incoming and outgoing photons.

These reactions have first been shown to obey scaling laws,their
energy dependence at fixed large angle  being described by the so-called
 {\it counting rules}~\cite{Bro73}.
These  pre-QCD studies are based on dimensional arguments and are not
specific of QCD. We will see later how they are realized (and slightly
modified)
in the framework of QCD. It is very instructive to first follow their
derivation
which leads to the correct physical picture of hard exclusive reactions.
\section{ Counting rules}

There are two standard ways to present them: a reasonning which emphasises the
space-time structure of these reactions and a dimensional
argument. They both are based on the hypothesis that the
elementary mechanism is hard, that is that {\it all} elementary
constituents  undergo a large momentum  transfer during
the short time process. A rigourous proof of the validity of this
hypothesis from field theory is not easy. We shall go back to this point.

\subsection{ Space-time picture; the example of the electromagnetic form
factor}

The simplest exclusive quantity is the {\it pion form factor }.
Let us  consider the process $e^-\pi^+\rightarrow e^-\pi^+$. The
electromagnetic interaction is mediated by virtual photon exchanges;
  effects due to the exchange of more than one
photon, of order $\alpha_{em}$ relatively to the exchange
of one photon, are negligible, and one thus limits the discussion to the
process
of Figure~1.

\vglue5cm
\centerline{\small Figure 1: The pion electromagnetic form factor}

\vskip1cm
The pion is a  pseudoscalar particule. If it were elementary,
the scattering cross section would equal
\beq
\left.{d\sigma\over dt}\right|_{\rm point}={4\pi\alpha^2\over t^2}
{(s-m^2-M^2)^2+t(s-m^2)\over\left(s-(m+M)^2\right)\left(s-(m-M)^2\right)},
\eeq
where $s=(k+p)^2$ and $t=(k-k')^2=-Q^2\le 0$. $m$ and $M$ are the
electron and pion masses.

The pion is however  composite and the cross section writes
\beq
{d\sigma\over dt}=|F_{\pi}(Q^2)|^2\,\left.{d\sigma\over dt}\right|_{\rm point}
\label{fpi}
\eeq
which defines the pion form factor $F_\pi$. It measures the ability of the
 pion to stay itself when being collided by an electron. It is thus a quantity
much sensitive to  confinement mechanisms. The physics deals with the
restauration of the meson integrity after the violent shock of a
 high-energy electron with one of the quarks. At the limit
$Q^2=0$, the meson  structure is not resolved, and $F_{\pi}(0)=1$.

The parametrization of Eq.(\ref{fpi}) is derived by writing
the matrix element~$S$ under the form
\beq
\langle e\pi|S|e\pi\rangle=\int d^4xd^4y
\langle\pi|J^{\mu}(x)|\pi\rangle\langle0|T(A_{\mu}(x)A_{\nu}(y))|0\rangle
\langle e|j^{\nu}(y)|e\rangle,
\eeq
where $J^{\mu}$ and $j^{\nu}$ are respectively quark and electron
electromagnetic currents. One thus  isolates the matrix element where
the pion structure plays a role
\beq
\langle\pi^+(p')|J^{\mu}(x)|\pi^+(p)\rangle=
\langle\pi^+(p')|J^{\mu}(0)|\pi^+(p)\rangle e^{i(p'-p)x}
\eeq

As the pion is a  (pseudo-)scalar particule, the most general  parametrization
of such a 4-vector must be written with the help of the 4-vectors
$(p+p')^{\mu}$ and $(p'-p)^{\mu}$ weighted by functions of $Q^2$, the only
scalar present in the problem (ignoring the pion mass, $m_{\pi}$). Since the
electromagnetic current is conserved: $\partial_{\mu}J^{\mu}=0$, the term in
 $(p'-p)^{\mu}$ must vanish. We thus have
\beq
\langle\pi^+(p')|J^{\mu}(0)|\pi^+(p)\rangle=e_{\pi}(p+p')^{\mu}F_{\pi}(Q^2).
\eeq

Note that the hermiticity of the current leads to the reality of the
form factor  (for a space-like transition).
\vskip1cm
Let us now derive the $Q^2$ dependence of the pion form factor $F_{\pi}(Q^2)$
(for large values of this variable) by a careful examination of the
way this process takes place. To do this, it turns out to be useful
to consider, in the  center of  mass frame of the reaction, the case,
illustrated
on  Figure~2, where the final electron emerges at an angle of 180$^{\circ}$
with respect to the initial electron.

\vglue5cm
\centerline{\small Figure 2: The space-time picture of the process
$e^-\pi^+\rightarrow e^-\pi^+$}
\vskip1cm

In its rest frame, the pion is represented as a
collection of partons,  quarks and gluons, {\it grosso-modo}
uniformously spreaded in a  sphere of  radius $R$ (typically the
pion charge radius, around $0.5\,$fm). In the reaction center of mass frame,
the longitudinal  dimension is Lorentz-contracted to $R/\gamma$ with
$\gamma=Q/2M$. The transverse dimensions are on the orher hand not
affected by Lorentz-contraction. At time 0, the electron hits one of the
 partons, the so-called {\it active} parton, and both change directions.
For the whole process to be elastic, {\it all} other partons
must be alerted before the moment $t\approx1/Q$ to
form the emerging pion  (also contracted in this frame).
The motion of the active parton after the collision is
$z(t)=-t,\ x(t),y(t)=0$ whereas the motion of a  {\it spectator} parton is
$z(t)=t+z_0,\ x(t)=x_0,\ y(t)=y_0$ (one has $-1/Q\lapprox z_0\lapprox 1/Q$ and
$-R\le x_0,y_0\le R$). Between the moments 0 and $1/Q$ a spectator parton
can receive and respond to a physical signal emitted by the active parton
at time 0
only if the interval $\Delta=t^2-(t+z_0)^2-x_0^2-y_0^2$ is positive, that is if
the spectator is at a distance $\sqrt{x_0^2+y_0^2}\lapprox1/Q$ in the
transverse
plane. One thus counts the probability to find spectator partons in a
transverse disc of  radius $1/Q$, in the initial as well as in the  final
state.
One gets
\beq
F_{\pi}^2\propto\left({\pi Q^{-2}\over \pi R^2_{\pi}}\right)
^{n_{in}-1+n_{out}-1}.
\eeq
Since a pion contains at least a valence quark and antiquark, we get a
minimal contribution scaling like $1/Q^2$. Adding for instance one
 gluon to the  valence in the initial state, without changing
the final state, yields a contribution scaling like $1/Q^3$ \dots\ These
contributions diminish relatively to the valence state contribution
 as energy increases.

This most important feature of the study of form factors at large transfer
 will be generalized to other exclusive reactions: {\it when the interaction is
at short distance, the valence  contributes in a  dominant way in terms of
scaling law}. Moreover, and this will be crucial for the phenomenon of
{\it color transparency}, the hadron configurations which contribute have
small ($O(1/Q)$) transverse sizes.

Let us summarize: asymptotically, one predicts for the energy dependence of
pion and nucleon form factors, a power-law fall off:
\beq
F_{\pi}(Q^2)\propto {1\over Q^2}~~~~~~~~~~~~~~
F_N(Q^2)\propto {1\over Q^4}.
\eeq
In the proton case, there are two form factors and the reasonning
developped here does not allow to distinguish them.
In fact, if one separates the form  factors with respect to their  degree
of helicity conservation,
one shows that the above counting rule applies only for helicity
conserving processes (and thus for the magnetic form factor $G_M$), but that
an additional power suppression affects $G_E$.

\subsection{Dimensional argument: the example of Compton scattering }

The dimensional argument to get the scaling law for exclusive processes
is quite general, but will be explained on the specific process
\beq
\gamma+N\rightarrow\gamma+N.
\eeq
In the ultra relativistic
limit,  the differential cross section writes
\beq
\label{sectionefficace}
{d\sigma\over dt}={1\over 16\pi s^2}|{\cal M}|^2,
\eeq
which we will consider for a fixed ratio $-t/s=O(1)$. To find the scaling
law of this reaction, we must identify the $s$ power dependence of the
amplitude ${\cal M}$.

This amplitude ${\cal M}$ is calculated from  Feynman rules and one may
 {\it a priori} identify the dimensions, in energy units, of
the different quantities entering these rules:
$$
\vbox{
\halign{#\tvi\hrulefill&\hfil#\cr
-- an external spinor  has a dimension &$1/2$,\cr
-- an external vector &0,\cr
-- a fermion propagator &$-1$,\cr
-- a  boson propagator &$-2$,\cr
-- a  boson-fermions vertex &0,\cr
-- a 3 gluon vertex &1,\cr
-- a 4 gluon vertex &0.\cr
}}$$

Let us now construct a  {\it tree level and connex} graph for the process at
the level of  elementary particules (see Fig.~3) and count the dimension
obtained: we get $-4$. One easily sees that this result does not depend
on an eventual insertion of additional loops; for
any connex graph, the dimension only depends on the number of  external
particles,
 $N$, and isequal to $4-N$.

\vglue5cm
\centerline{\small Figure 3: A connex graph contributing to Compton scattering}

When calculating ${\cal M}$, one must find out the momenta carried by each
line and compute the scalar products between these various momenta. One easily
 sees that distributing to each quark or gluon a {\it finite} fraction of
its parent hadron momentum, leads all particules to undergo a large momentum
transfer (if, of course, the global transfer is sufficiently large). Then
all scalar products are of order $s$, which is the unique dimensionful
scale in the kinematics studied.

\smallskip
In  the conventions where spinors are normalized by $\bar u~u~= 2m$, the
overall
 dimension of ${\cal M}$ vanishes.
We indeed must add a rule to the above list to precise how a hadron exhibits
its quark - gluon content and to quantify the transition from the hadron to a
 $n$ parton system. This transition,
\beq
|\hbox{Hadron}\rangle\leftrightarrow f_{H,n}|n\hbox{ partons}\rangle,
\eeq
introduces a constante $f_{H,n}$ of dimension $n-1$ which should be
independent of the particular hard reaction studied and
comes from  confinement physics. The natural energy scale\footnote{$M$
will in the following represent  a low energy scale, which can be
the  QCD constant, $\Lambda_{QCD}$, the $\rho$ meson mass or a typical
internal transverse momentum for meson constituents,
$\sqrt{\langle{\bf k}_{\bot}^2\rangle}$, {\it i.e.} a few hundreds of MeV.},
$M$, for the constants $f_{H,n}$ should thus be $s$-independent.
Taking  these  hadron-partons transitions into account, we find
\beq
{\cal M}=f_{H,n}\,\sqrt{s}^{\,4-n-1-n'-1}f_{H,n'},
\eeq
which is  dimensionless as it should be.

The large angle and high energy behaviour of the Compton differential
can thus be written as
\beq
{d\sigma\over dt}\sim {1\over  s^6} f({t\over s}),
\eeq
for the transition between {\it  valence} states.  The above study indicates
 indeed that the sub-process
$qqqg\gamma\rightarrow qqq\gamma$ contributes to the cross section as
\beq
{f^2_{N,{\rm val}+g}f^2_{N,{\rm val}}\over s^7},
\eeq
which is  negligible at high energy.

The amplitude at large transfer is thus separated as
\beq
A(s,t/s)=A^{\rm LT}\left(1+O(M/\sqrt{s})\right),\ t/s=O(1)
\eeq
where the so-called ``Leading Twist'' contribution $A^{\rm LT}$, yields the
lowest power fall-off in $s^{-1}$. The QCD analysis presented in section 2 will
strengthen the argument presented here and develop a consistent way of
calculating
the leading contribution. It will however be important to phenomenologically
verify that the scaling laws, and thus the dominance of valence states, are
verified at accessible energies, and this for each physical process under
study.

Before studying in more details the pedagogical case of the electromagnetic
meson form factor at large $Q^2$, let us digress to an important exception
to the counting rules, the so-called ``Landshoff'' process of multiple or
independent scattering.

\subsection{The exceptional case of  independent scattering}
This independent scattering process~\cite{Lan} does not appear in
electromagnetic
form factors but in elastic scattering of hadrons; it  is
represented on Figure 4 for the case of $\pi$-$\pi$ elastic scattering.
The power counting of this process goes as follows. The outgoing beams of
quarks
must coincide in direction well enough to make hadrons in the final state; any
discrepancy is set by the wave functions, which are defined\footnote{In
perturbation theory it is necessary to separate bound state properties of the
wave function from effects of gluon exchange. To avoid double counting the
gluon exchange which produces large $k_T$, the bound state wave functions
should have large $k_T$ tails subtracted.}  to have small relative $k_T$.
The allowed $k_T$ values are much smaller than the beam energies, so we
can approximate them as almost zero. Because each independent on--shell
quark-quark scattering amplitude scales like $g^2 \bar u u  \bar u u /t$
the independent scattering matrix element scales like
\beq
[g^2 \bar u u  \bar u u /t ]^{n/4} \sim g^{n/2}
\eeq
up to logarithmic corrections.The scaling behaviour
of the desired elastic scattering cross section comes then from the
integration region constraint on 4-momenta set by:
$\delta^4(k^1+k^2-k^3-k^4)$. There are three large momenta
for each scattering, and one out-of plane transverse momentum.  This component
of the transverse momentum is not as big as $\sqrt{s}$ but instead depends
sensitively on what the hadronic wave function  allows.  It should be of
order $C<k_T^2>^{1/2}$  in the state's wave function, which for purposes of
counting is the same as $C/<b^2>^{1/2}, b$ being a transverse space separation.

\vglue5cm
\centerline{\small Figure 4: The Independent Scattering  process }

Each delta function of a big momentum counts as $1/\sqrt{s}$, since
\beq
 \delta(p-p') \sim s^{-1/2}\delta(x-x'),
\eeq
 where $x$ and $x'$ are dimensionless scaling variables.
The overall probability amplitude  for a pair of quarks to coincide
in final state direction to make a hadron scales like the product of the delta
functions of momentum, namely like $C<b^2>^{1/2} (s)^{-3/2}$. Using
Eq.~(\ref{sectionefficace}), one finds
\beq
{d\sigma\over dt}\propto <b^2> s^{-5},
\eeq
for meson-meson scattering. As $s\rightarrow \infty$, this beats the
quark-counting process, which for meson-meson scattering goes like $s^{-6}$.

 Consider next proton-proton scattering.   The argument goes the same way, but
requires another quark-quark scattering to coincide with the first ones.
This adds three more delta functions of big momenta, so the amplitude-squared
is smaller by $s^{-3}$.  Independent $pp\longrightarrow pp$ scattering thus has
\beq
{d\sigma\over dt}\propto (<b^2>)^2 s^{-8}.
\eeq
  This again beats the quark-counting process, which (recall)
goes like $s^{-10}$.

 How did this process manage to evade the power counting of the quark counting
process?  It is easy to show that the number of gluons and internal propagators
is fewer than the one assumed in the quark-counting induction; the topologies
of the low order diagrams are not the same.
Because {\it both} quark counting and independent scattering were studied
before QCD was established, early discussion focused on comparisons with data.
At first it seemed as if $p$-$p$ scattering went like $s^{-10}$, creating a
puzzle to explain the absence of the much bigger $s^{-8}$. One argument was
made that quark counting diagrams might be more numerous and would dominate for
that reason. However, when
compared at the same order of perturbation theory, the independent scattering
graphs are myriad and re-emerge inside the quark counting diagrams.
This happens because internal gluons can become ``soft": a diagram with a
soft gluon scales with the same power of $s$ as if this gluon was absent.
The upshot is that many quark counting
diagrams contain a region indistinguishable from independent
scattering with a soft gluon. Independent scattering cannot in any sense be
``absent". Similarly, if ``soft" gluons attached to an independent scattering
diagram should receive enough momentum to be counted as ``hard", the diagram
may merge into the quark--counting set. It was finally realized \cite{BS89}
 that these physically distinct processes actually boil down to different
integration regions found in the one theory of QCD.

      The independent scattering process had a confused history as these
subtleties were only gradually appreciated.  Closely related (and as much
confused) is the issue of ``Sudakov effects", at first thought to
suppress the independent scattering regions, but which were subsequently shown
actually to force the independent scattering to dominate in the limit of
$s\longrightarrow
\infty$. We believe that there is rather convincing evidence that independent
scattering region of QCD has
been observed and plays a major role in color transparency.
However, the subject is unsettled, and the interplay of the independent
scattering regions and the quark counting regions is currently a subject of
active investigation.

\section{ Calculating the  pion form factor }

One now wants to really calculate from QCD the pion form factor at large
transfer
\cite{Far79}. This leads us to precise first the hadron wave function and the
Born hard amplitudes, then the radiative corrections to see if a sensible
picture
emerges where a non perturbative object sensitive to confinement dynamics
factorizes
from a hard scattering amplitude  controlled by a perturbative expansion
which is renormalization group improved. This factorization which is
crucial for a
consistent understanding of future experimental data may be pictiorally
described
as in Fig.5.

\vglue5cm
\centerline{\small Figure 5: Factorization of a hard exclusive process :
$X*T_H*X'$}
We restrict here to the pedagogical case of the $\pi$  meson form factor but
the
technique is applicable to all hard exclusive reactions.

\subsection{ Description of the pion}

Let us specify the kinematics. In the Breit frame
the momenta are written as:
\begin{equation}
q=\pmatrix{0\cr0\cr0\cr Q\cr},\
p=\pmatrix{Q/2\cr0\cr0\cr -Q/2\cr},\
p'=\pmatrix{Q/2\cr0\cr0\cr Q/2\cr};
\end{equation}
where the pion energies
$$E_{\pi}={Q\over2}\left(\sqrt{1+{4m_{\pi}^2\over Q^2}}\right)$$
have been approximated by  $Q/2$.

To describe the  pion  in its  valence state, one introduces  the
Bethe-Salpeter~(BS) amplitude~\cite{Sal51}
\begin{equation}
\langle0|T\left(q_{u\alpha i}(y)~P_{ij}(y,0)~\bar{q}_{d\beta
j}(0)\right)|\pi^+(p)\rangle,
\end{equation}
where $u$ and $\bar{d}$ are the flavours of  the valence quarks of $\pi^+$,
$\alpha$ and $\beta$ are Dirac indices and $i$, $j$ are color
indices. The $P_{ij}$ operator is necessary to have an amplitude invariant
under
local gauge transformations; when $q(y)$ transforms to $U(y)~q(y)$, $P(y,0)$
transforms to $ U^{-1}(y)~P(y,0)~U(0) $, compensating the quark and antiquark
 variations. The BS amplitude is the relativistic generalisation of the
 Shr\"odinger wave function  describing the bound state
of a quark antiquark pair ~\cite{Lurie}. One may interpret it as the
 probability amplitude of finding in a $\pi^+$ a $u$ quark at
point $y$ and a  $\bar{d}$ antiquark at the origin.

One often prefers to work in momentum space and defines the Fourier
 transform of the BS amplitude as
\begin{equation}
\int d^4ye^{ik.y}<>=X_{\alpha\beta}(k,p-k)
\end{equation}
where $k$ is the quark momentum and, by momentum conservation,
$p-k$ is the antiquark momentum.

To discuss the properties of this amplitude, it is convenient to
introduce light-cone coordinates defined as:
\begin{equation}
\left\{\matrix{
k^+={1\over \sqrt{2}}(k^0-k^3)\cr
k^-={1\over \sqrt{2}}(k^0+k^3)\cr}\right.
\end{equation}
The  scalar product of two $4-$vectors A and B is then
\begin{equation}
A.B=A^+B^- +A^-B^+ -{\bf A}_{\bot}.{\bf B}_{\bot}.
\end{equation}
In our case, we thus have (listing $p=[p^+,p^-,p^1,p^2]$)
\begin{equation}
p=[Q/\sqrt{2},0,0,0],\ p'=[0,Q/\sqrt{2},0,0],
\end{equation}
and we parametrize the internal momenta as
$k=[xQ/\sqrt{2},k^-,{\bf k}_{\bot}]$,
where $x$ is the light-cone fraction carried by the  quark inside the pion.
The antiquark then carries the fraction $1-x=\bar{x}$. The final  pion is
treated similarly, with $+$ and $-$ components exchanged:
$k'=[k'^+,x'Q/\sqrt{2},{\bf k'}_{\bot}]$ and so on.

In terms of these  variables, the $k^{\mu}$ regions  favored by the
amplitude $X(k,p-k)$ are simply written as:
\begin{equation}
{\bf k}_{\bot}^2\lapprox M^2,\hskip 1cm |k^-|\lapprox M^2/Q.
\end{equation}


\subsection{The hard scattering at the Born level}

The matrix element of  Figure~5 is written as the convolution
\begin{equation}
\int {d^4k\over (2\pi)^4}{d^4k'\over (2\pi)^4}\,
X(k)\,T_H^{\mu}(k,k')\,X^{\dag}(k').
\end{equation}
At the lowest order in the QCD coupling constant, $g$, one finds 4
Feynman diagrams. One is drawn on   Figure~6 and the 3 others are easily
deduced
by attaching successively the photon to the  points 2, 3and 4.

Let us first evaluate the gluon squared momentum, which is in Feynman gauge,
the denominator of the gluon propagator. We have
\begin{equation}
\label{virtualite}
\matrix{
(p'-k'-p+k)^2=&-\bar{x}\bar{x}'Q^2
&-\sqrt{2}Q(k^-\bar{x}'+k'^-\bar{x})&-2k^-k'^+
&-({\bf k}_{\bot}-{\bf k}'_{\bot})^2\cr
\noalign{\medskip}
&O(Q^2)&O(M^2)&O({\displaystyle M^4\over\displaystyle Q^2})&O(M^2)\cr
}
\end{equation}
where typical orders of magnitude indicated refer to the momentum regions
favored by the amplitudes $X(k)$ and $X(k')$. Restricting to leading terms in
$Q$,we may forget terms of order $M^2$. So, in particular, we write
\begin{equation}
\label{virtualiteapprox}
(p'-k'-p+k)^2\approx-\bar{x}\bar{x}'{\displaystyle Q^2\over\displaystyle 2}.
\end{equation}

The same analysis may be repeated for the other quantities present
in the hard amplitude  $T_H^{\mu}$, leading to
\begin{equation}
T_H^{\mu}(k,k')\approx
T_H^{\mu}\left(x{Q\over\sqrt{2}},x'{Q\over\sqrt{2}}\right).
\end{equation}

\vglue5cm
{\center \small Figure 6: Born Graph for the pion form  factor; the 3 other
graphs
are deduced by attaching the photon to the  points 2, 3 and 4. Propagators
joining   Bethe-Salpeter amplitudes to the vertices are
absorbed, by definition, in these amplitudes.}

\vskip1cm
We may then express the convolution of equation~(30) under the
form
\begin{equation}
\int dxdx'\,
\left({Q\over2\sqrt{2}\pi}\int {dk^-d{\bf k}_{\bot}\over (2\pi)^3}X(k)\right)
T_H^{\mu}(x,x')
\left({Q\over2\sqrt{2}\pi}
\int {dk'^+d{\bf k'}_{\bot}\over (2\pi)^3}X^{\dag}(k')\right),
\end{equation}
and the object needed to describe the pion in this reaction is in fact much
simpler than the amplitude $X$ since one may integrate over three components
of the internal momentum.

A first simplification comes from the integration over the $k^-$ (for the
 outgoing pion over the $k'^+$ variable).
In  terms of the conjugated variable $y^+$, this means that one only needs the
 Bethe-Salpeter amplitude at $y^+=0$, which is called the light cone wave
 function , usually noted as  $\psi(x,{\bf k}_{\bot})$~\cite{Bro}. A
useful property of this wave function is that the support in the  $x$
fraction is limited, as $0\le x\le 1$.
This limitation to light cone fractions  $x$ between  0 and 1
may be recovered by writing $X$ under the form

\begin{equation}
X(k,p-k)={f(k)\over [k^2-m^2+i\varepsilon]\ [(p-k)^2-m^2+i\varepsilon]},
\end{equation}
and by evaluating the integral over  $k^-$ from $-\infty$ to $+\infty$
by a Cauchy contour. One then obtains a non zero contribution only if the two
 poles are on opposite sides of the real axis. These
poles are at
\begin{equation}
\left\{\matrix{
k_1^-&=&{\sqrt{2}({\bf k}_{\bot}^2+m^2)\over xQ}
&-i\varepsilon\,{\rm sgn}(x)\hfill\cr
k_2^-&=&-{\sqrt{2}({\bf k}_{\bot}^2+m^2)\over \bar{x}Q}
&+i\varepsilon\,{\rm sgn}(1-x)\cr
}\right.
\end{equation}
so that the integral yields a factor\footnote{
$\theta$ is the  function defined  by :
$\theta(x)=\left\{\matrix{0,&x<0,\cr 1,& x>0.\cr}.\right.$
}
\begin{equation}
\theta(x)\theta(1-x);
\end{equation}
the $x$ integral is thus limited to the interval $[\,0,\,1\,]$.

The Dirac structure of the amplitude $X(k)$ integrated over $k^-$ and
${\bf k}_{\bot}$ is easy to extract~\cite{Gou} and one finds
\beq
M_{\alpha\beta}(x,p)=
{1\over4}\gamma^5p\slash\,\varphi(x)\big|_{\alpha\beta}.
\eeq
This Dirac structure corresponds to the combination of spinors
($\uparrow$ and $\downarrow$ denote respectively the helicity states $+$
and $-$)
\beq
{1\over4}\gamma^5\,p\slash|_{\alpha\beta}=
{1\over2\sqrt{2x\bar{x}}}{1\over\sqrt{2}}\left(
u_{\alpha}(x{\bf p},\uparrow)\,\bar{v}(\bar{x}{\bf p},\downarrow)
-u_{\alpha}(x{\bf p},\downarrow)\,\bar{v}(\bar{x}{\bf p},\uparrow)\right),
\eeq
{\it i.e.} one recovers the pion spin wave function in the quark  model
${1\over\sqrt{2}}\left(|\uparrow\downarrow\rangle
-|\downarrow\uparrow\rangle\right)$.

The function $\varphi(x)$ is called {\it distribution amplitude}; it
``measures'' how the  pion momentum is  distributed between the valence
quark and
antiquark when their transverse separationvanishes. {\it This is the
 non perturbative amplitude  connecting long distance physics
of strong interaction to short distance hard processes}.

Let us now precise a little bit the color algebra involved here. A useful
way to
simplify this matter is to choose for a pion of momentum along the $+$
direction,
axial gauges with axis along the  $-$ direction ( fixing
$A^+=0$)~\cite{Gou}. In
these gauges, one has $P_{ij}(y,0)=\delta_{ij}$ and the color component for the
quark-antiquark pair is simply $\delta_{ij}/3$.
This fact partly explains the interest of light-cone gauges in the study
of hard processes. For another gauge choice, an explicit form of $P_{ij}(y,0)$
is necessary, but we will not pursue this here. Note however that $P_{ij}(y,0)$
may be perturbatively analyzed and gauge invariance preserved order by
order in the perturbative expansion. At zeroth order, one has
\beq
P_{ij}(y,0)=\delta_{ij}+O(g).
\eeq

We are now able to calculate the graph of Figure~5 with a new Feynman rule
for the pion
\beq
{1\over 3}\delta_{ij}{1\over4}\gamma^5\,p\slash|_{\alpha\beta}\,\varphi(x),
\eeq
and a loop integral $\int_0^1dx$. The amplitude of the  process may thus be
written as
\begin{equation}
\int_0^1dx\int_0^1dx'\varphi(x)\langle T_H^{\mu}(x,x')\rangle\varphi^*(x')
\end{equation}
where the hard process is evaluated on the spin and color components written
above. Color algebra leads to the trace
\begin{equation}
{1\over3}\delta_{ij}T^a_{jk}{1\over 3}{\delta_{kl}\over3}T^a_{li}
={C_F\over3}={4\over 9},
\end{equation}
and the amplitude  neglecting quark masses is
\begin{eqnarray}
&&\int_0^1dx\int_0^1dx'
(-){C_F\over3}Tr\left\{e_u\gamma^{\mu}{1\over4}\gamma^5p\slash\,
g\gamma^{\alpha}{1\over4}\gamma^5p\slash'\,g\gamma^{\beta}
{p'-\bar{x}p\over -\bar{x}Q^2}\right\}
{-\eta_{\alpha\beta}\over -\bar{x}\bar{x}'Q^2}\varphi(x)\varphi^*(x')
\nonumber\\
&=& e_up^{\mu}{C_Fg^2\over6Q^2}
\left|\int_0^1dx{\varphi(x)\over\bar{x}}\right|^2.
\end{eqnarray}

The graph with the photon  attached to  point 2 leads to the same
expression replacing $p^{\mu}$ by $p'^{\mu}$. The two other graphs
are identical to the two first ones after exchanging $e_u\leftrightarrow-e_d$
and $\bar{x}\leftrightarrow x$ in the integrand denominator.
Charge  conjugation invariance and  isospin symmetry lead to the
relation $\varphi(x)=\varphi(\bar{x})$, so that one can factorize the
term $(e_u-e_d)(p+p')^{\mu}$ expected in Eq.~(8) and
isolate the form factor expression
\begin{equation}
F_{\pi}(Q^2)=
{C_Fg^2\over6Q^2}\left|\int_0^1dx{\varphi(x)\over\bar{x}}\right|^2.
\end{equation}

Let us stress that we recover the scaling law in  $Q^{-2}$ predicted
by the counting rules.
\smallskip

The pion lifetime fixes a constraint on the valence wave function of the pion.
 The process is decribed on Figure~7.

\vglue5cm
\centerline{\small Figure 7: pion weak decay.}

As in the form factor case, one may isolate the weak transition at the quark
level, under the form of the matrix element of the electroweak current
\cite{Donoghue}. One gets
\begin{equation}
\langle0|\bar{q}_d(0)\gamma^{\mu}(1-\gamma^5)q_u(0)|\pi^+(p)\rangle
=f_{\pi}p^{\mu},
\end{equation}
where the decay constant, $f_{\pi}$, is in this parametrization equal to
133MeV.

The BS amplitude at the origin may then be written as
\begin{equation}
\langle0|T\left(q_{u\alpha i}(0)\bar{q}_{d\beta j}(0)\right)|\pi^+(p)\rangle
=\int_0^1dx{Q\over2\sqrt{2}\pi}\int {dk^-d{\bf k}_{\bot}\over (2\pi)^3}X(k),
\end{equation}
that one multiplies by the tensor
$[\gamma^{\mu}(1-\gamma^5)]_{\beta\alpha}\delta_{ji}$ to get
\begin{equation}
-\langle0|\bar{q}_{d i}(0)\gamma^{\mu}(1-\gamma^5)
q_{u i}(0)|\pi^+(p)\rangle
=Tr\left({\gamma^5p\slash\over4}\gamma^{\mu}(1-\gamma^5)\right)
{\delta_{ij}\over3}\delta_{ji}\int_0^1dx\,\varphi(x),
\end{equation}
where it can be noted that the componant $\varphi'$ does not  survive to the
projection. One gets
\begin{equation}
p^{\mu}\int_0^1dx\,\varphi(x)=f_{\pi}p^{\mu},
\end{equation}
which fixes the normalization of the distribution amplitude.

\subsection{ Radiative corrections}

It is important, when calculating a quantity in any field theory, and in
particular in  perturbative QCD, to keep track of  radiative corrections
and control them so that the picture obtained at lowest order survives their
inclusion.
The ultraviolet regimes does not a priori cause much problem since the
 theory is known to be  renormalizable. In fact, the subtractions
to be taken into account  are automatically taken care of when
 correctly treating quark and gluon propagators on the one hand,
and the running coupling constant on the other hand.

The infrared regions in the loop calculations must be very carefully
scrutinized.
In the specific process studied here, one finds in a  $n$ loops diagram
corrections of order
\beq
{\alpha_S(Q^2)\over Q^2}\left[\alpha_S(Q^2)\ln {Q^2\over M^2}\right]^n,
\eeq
which, since $\alpha_S(Q^2)\propto(\ln Q^2/\Lambda^2)^{-1}$ is of
the same order as the tree level process! One has to resum these large
logarithms in the distribution amplitude to recover the predictibility
of the  formalism. This is {\it factorization} since then the process may
be written as the convolution illustrated by Figure~5:
\beq
F_{\pi}=\varphi*T*\varphi^*
\eeq
where:

-- $T$ is a hard amplitude that one can evaluate within
 perturbative  QCD; namely, higher order corrections to $T$ are of order
 $\alpha_S^n(Q)$, and thus sufficiently small at sufficiently
large transfer;

-- all large  logarithms are absorbed in $\varphi$; the
distribution $\varphi$, which represents the wave function evolves
with the scale $Q$ characteristic of the virtual photon probe.
This stays an essentially non perturbative quantity expressing the way
confined valence quarks share the hadron momentum when they interact at
small distance in an exclusive process.

Let us now examine how leading logarithms are resummed in the distribution
 $\varphi_{\rm LL}$~\cite{Field}. It turns out that it is most interesting to
choose to work in a gauge which is different from the Feynman gauge, namely
an axial gauge, with axis $n^{\mu}$,  fixing the  condition on gluon fields
 $A^{\mu}$ as: $n.A=0$. The leading  corrections have then the
 form illustrated on Figure~8.

One may show that the graph summation  yields
\beqn
\varphi_{\rm LL}(x,Q)&=&\,\varphi_0(x)
+\kappa\int_0^1du\,V_{q\bar{q}\rightarrow
q\bar{q}}(u,x)\,\varphi_0(u)\nonumber\\
&+&{\kappa^2\over2!}\int_0^1duV_{q\bar{q}\rightarrow q\bar{q}}(u,x)
\int_0^1du'V_{q\bar{q}\rightarrow q\bar{q}}(u',u)\,\varphi_0(u')+\ldots
\eeqn
where
$\kappa$ contains large collinear logarithms under the form
\beq
\kappa={1\over\beta_1}\ln{\alpha_S(\mu^2)\over\alpha_S(Q^2)}\ \
\left(\beta_1={1\over 4}\left(11-{2\over3}n_f\right)\right);
\eeq
and $V_{q\bar{q}\rightarrow q\bar{q}}$ is a characteristic kernel describing
the splitting of the
 valence distribution of the pion
\beq
V_{q\bar{q}\rightarrow q\bar{q}}(u,x)={2\over3}\left\{
{\bar{x}\over\bar{u}}\left(1+{1\over u-x}\right)_+\theta(u-x)+
{x\over u}\left(1+{1\over x-u}\right)_+\theta(x-u)\right\},
\eeq
The $()_+$ distribution comes from the compensation of infrared divergences
 (here in the limit $u\rightarrow x$)  between
 graphs b and c of  Figure~8. This is a consequence
of the colour neutrality of a hadron.

\vglue5cm
\centerline{\small Figure 8: Leading corrections in axial gauge}

The equation on $\varphi$ (omitting from now on the index LL) may be rewritten
under the integro-differential form
\beq
\left({\partial\varphi\over\partial\kappa}\right)_x=
\int_0^1du\,V(u,x)\,\varphi(u,Q),
\eeq
the general solution of which is  known as
\beq
\varphi(x,Q)=x(1-x)\sum_n\phi_n(Q)C_n^{(3/2)}(2x-1);
\eeq
where Gegenbauer polynomials   $C_n^{(m)}$ are such that
\beq
\int_0^1du\,u(1-u)\,V(u,x)\,C_n^{(3/2)}(2u-1)=A_nx(1-x)C_n^{(3/2)}(2x-1),
\eeq
with $A_n$  coefficients which depend on $n$. Injecting this
solution in the equation, one gets
\beq
\phi_n(Q)=\phi_n(\mu)e^{A_n\kappa}=\phi_n(\mu)\left(
{\alpha_S(\mu^2)\over\alpha_S(Q^2)}\right)^{A_n/\beta_1},
\eeq
where the  exponents in the expansion begin with
\beq
{A_0\over\beta_1}=0,\ {A_2\over\beta_1}=-0,62,\ \ldots
\eeq
Odd terms disappear since the distribution is symmetric in the interval
$[\,0,\,1\,]$.

Calculating the integral
\beq
\int_0^1dx\,\varphi(x,Q)=\phi_0(Q)\int_0^1dx\,x(1-x)={\phi_0\over6}=f_{\pi}
\eeq
one can write down the beginning of the expansion:
\beq
\varphi(x,Q)=6f_{\pi}x(1-x)+(\ln Q^2)^{-0.62}\,\Phi_2\,x(1-x)[5(2x-1)^2-1]
+\ldots
\eeq
The pion asymptotic distribution, when $Q\rightarrow\infty$,  is then
\beq
\varphi(x,Q\rightarrow\infty)\sim 6f_{\pi}x(1-x).
\eeq
This however does not tell us much on the realistic distribution amplitude
at accessible energies: the constants $\Phi_2,\ldots\Phi_n$ are unknown.

This is how far perturbative QCD can lead us about
the distribution amplitude $\varphi$; {\it i.e.} to understand how
strong interactions build a hadron from its valence quarks. To go further, one
needs other methods, which are non perturbative by nature. Experiments can
guide us to
develop new ways since exclusive scattering data may be processed to extract
distribution amplitudes. The existing methods, like lattice calculations or QCD
sum rules, are still too primitive and rely on too many unchecked hypotheses to
be trusted. They however lead to useful rate estimates. They generally
evaluate {\it moments} of the distribution amplitude defined as:
\beq
\int_0^1dx(2x-1)^2\varphi(x,\mu),\ \ldots
\eeq
Such a study lead Chernyak and Zhitnitsky~\cite{Che84} to propose the
distribution
\beq
\varphi_{\rm cz}(x,Q^2)=6f_{\pi}x(1-x)\left\{1+[5(2x-1)^2-1]\left(
{\ln Q^2/\Lambda^2\over\ln Q_0^2/\Lambda^2}\right)^{-0.62}\right\},
\eeq
with $Q_0\approx500$MeV.  Figure~9  shows the distribution proposed by Chernyak
 and Zhitnitski.

\vglue5cm
\centerline{\small Figure 9: The CZ distribution and its
evolution with the scale $\mu^2$.}

\subsection{ Transverse Degrees of Freedom}

A study of one loop corrections~\cite{Dit81} leads to propose that
 the scale relevant for the running coupling constant $\alpha_S$
is more likely to be the exchanged gluon  virtuality $xx'Q^2$ than the
photon virtuality $Q^2$.  The whole treatment would then be correct
only when the gluon is far off mass shell, that is as far as $x$ or $x'$ do
not approach 0.  However, for intermediate transfers, it turns out that
an important part of the amplitude comes from these regions. One should
thus reexamin the whole story in the region where gluons may become soft.
In this region transverse momentum (or transverse distance) degrees of freedom
become important and invalidate the collinear approximation~\cite{Li92}.
Let us qualitatively explain the expected  modifications.

 The elastic interaction of a coloured object (a quark for instance) is
suppressed by a Sudakov form factor~\cite{Sud56} which quantifies the
difficulty
 of preventing an accelerated charge  from  radiating.
Similarly the elastic interaction of a dipole of  transverse size $b$
is strongly suppressed unless $b$ approaches $Q^{-1}$~\cite{BS89}.
The approximation where transverse  degrees of freedom are
 neglected leads to consider the region
$b^2 \leq (xx'|q^2|)^{-1}$, which is an unsuppressed region
when $xx'$ is of order 1. When $xx'\rightarrow 0$, this
approximation becomes illegitimate, and one may imagine that taking
  the transverse size into account should allow, with the help of an
 associated Sudakov suppression, to bypass dangerous
infrared contribution. We shall come back to this point in section 4.2.

\section{ Other scattering processes. }

The results obtained above for the  electromagnetic form factor may be
generalized to other hard exclusive  processes, with an important difference
in the case of  hadron - hadron collisions  (see section 4). One thus
defines a distribution amplitude for the proton and analyzes the magnetic
form factor $G_M$ very similarly. One can then  consider  sharper
 reactions as real or virtual  Compton scattering, which still only depend on
the proton structure but where one can vary dimensionless ratio such as angles.

\subsection{The proton distribution amplitude}

As for the  pion case, the  valence nucleon wave function can be
written~\cite{Gou}
as a combinationof definite tensors of colour, flavour and spinor indices with
a
(unique) proton distribution amplitude $\varphi(x,y,z)$.
This distribution amplitude may be written as an expansion quite similar to
what
was derived above for the pion case but on a different basis of polynomials:
\beqn
\varphi(x_i,Q)= &120 x_1 x_2 x_3 ~\delta(x_1+x_2+x_3-1) \times \nonumber\\
 & \left[1 + {{21}\over{2}}
\left({\alpha_S(Q^2)\over\alpha_S(Q_0^2)}\right)^{\lambda_1} A_1
 P_1(x_i)+
{{7}\over{2}} \left({\alpha_S(Q^2)\over\alpha_S(Q_0^2)}\right)^{\lambda_2}
 A_2 P_2(x_i) + \dots \right] ,
\eeqn
where the slow $Q^2$ evolution comes entirely from the terms
$ \alpha_S(Q^2)^{\lambda_i}$, and the  $ \lambda_i$'s are decreasing numbers:
\beq
\lambda_1 = {{5}\over{9\beta_1}}   ~~~~,~~~~\lambda_2 = {{6}\over {9\beta_1}} ,
\eeq
and the $P_i(x_j) $'s are Appell polynomials:
\beq
P_1(x_i)=x_1-x_3 ~~~,~~~ P_2(x_i)=1-3x_2 ,...
\eeq
The $A_i$'s are unknown constants and measure the wave function projection on
the Appell polynomials:
\beq
A_i=\int _{0}^{1}dx_1 dx_2 dx_3 ~\delta(x_1+x_2+x_3-1)~\varphi(x_i)~P_i(x_i)
\label{Appell}
\eeq

\subsection{The proton magnetic  form factor}

One  describes the elastic interaction
of a proton and an electron
\beq
e^- +p\rightarrow e^- +p,
\eeq
with two form factors $F_1$ and $F_2$ (still within the one virtual photon
exchange hypothesis)
\beq
\langle p',h'|J^{\mu}(0)|p,h\rangle=e\bar{u}(p',h')\left[
F_1(Q^2)\gamma^{\mu}+i{\kappa\over2M}F_2(Q^2)\sigma^{\mu\nu}(p'-p)_{\nu}
\right]u(p,h);
\eeq
$h$ and $h'$ are respectively the incoming and outgoing proton helicities,
 $u$ and $\bar{u}$ their spinors and  $M$ the
proton mass. In this decomposition, $e$ is the proton charge and
 $\kappa=1.79$ is its anomalous magnetic moment.
With these conventions, the two form factors have at zero transfer the values:
\beq
F_1(0)=1,\hskip3cm F_2(0)=1.
\eeq

\vglue5cm
{\centerline {\small Figure 10: $Q^2$ evolution of the proton magnetic form
 factor~\cite{Sil93}}}

 $F_1$ and $F_2$ are respectively called
 Dirac and Pauli form factors. From the Gordon identity
\beq
i(p'-p)_{\nu}\bar{u}'\sigma^{\mu\nu}u=
2M\bar{u}'\gamma^{\mu} u-(p+p')^{\mu}\bar{u}'u,
\eeq
one writes the current matrix element as
\beq
\langle p',h'|J^{\mu}(0)|p,h\rangle=e\bar{u}'\left[
(F_1(Q^2)+\kappa F_2(Q^2))\gamma^{\mu}-{\kappa\over2M}F_2(Q^2)(p+p')^{\mu}
\right]u,
\eeq
which leads to define the Sachs form factors which appear in the
 process cross section; they are the linear combinations
\beqn
G_M&=&F_1+\kappa F_2\nonumber\\
G_E&=&F_1+{q^2\over 4M^2}\kappa F_2.
\eeqn

In the  formalism we are presenting here, only the magnetic form
factor is accessible. With a proton distribution amplitude deduced from
a QCD sum rule analysis \`a la Chernyak-Zhitnitsky~\cite{Che84}, one
obtains the
results shown on Figure~10. The slight decrease of $Q^4 G_M(Q^2)$ is understood
as a manifestation of radiative corrections on top of the counting rules.

\subsection{ Compton scattering}

The  perturbative part of the analysis of real~\cite{Far88,Kro91} or
virtual~\cite{Far91} Compton scattering consists in evaluating the 336
topologically distinct diagrams obtained when coupling two photons to the
three valence quarks of the proton, two gluons being exchanged. Moreover,
there are
 42 diagrams with a three-gluon coupling but it turns out that their color
factor vanishes.

At lowest order in  $\alpha \sim {{1} \over {137}}$, Virtual Compton
Scattering  (VCS) is described as the  coherent sum of the amplitudes
drawn on  Figure 11, namely the Bethe Heitler (BH) process (Fig. 11b)
where the  final photon is radiated from the electron and the genuine
VCS process of (Fig. 11a).

\vglue5cm
\centerline{\small Figure 11: Virtual Compton Scattering amplitudes}

As the BH amplitude is calculable from the elastic form
factors $G_{Mp}(Q^2)$ and $G_{Ep}(Q^2)$,
 its interference with the VCS amplitude is an  interesting source of
information, different from what real
Compton scattering yields. The VCS amplitude depends on three invariants;
 one usually chooses $ Q^2,  s, t $ or $s, Q^2/s, \theta_{CM} $.

Each incoming (outgoing) quark carries a light-cone fraction $x$
($y$) of the $+$ ($-$) component of the parent proton momentum,
together with  components along the three other directions. When these
fractions $x$ or $y$ stay of order 1, it is legitimate to  neglect these
three other components in the hard process and one gets:
\beq
A=\varphi_{(uud)}\otimes T_H(\{x\},\{y\})\otimes \varphi'_{(uud)}(1+O(M^2/t)),
\eeq

\vglue5cm
\centerline{\small Figure 12: Real  Compton Scattering at large angle}
\vskip1cm

Figure 12 shows the few existing data for real Compton scattering on the proton
with $-t>1GeV^2$ \cite{Shupe}. $s^6 d\sigma/dt$ is plotted as a function
of  $\cos\theta_{CM}$ to illustrate the approach to asymptotic scaling laws.
If one fits the data with a law in  $s^{-\alpha}$, one gets $\alpha=
7.0\pm0.4$:
that is a  $ 2.5\,\sigma$ deviation from the counting rule  prediction
$\alpha=6$.

\subsection{ A strategy for data analysis }

A first way to extract physics from experimental points consists in comparing
data with a computation done with distribution amplitudes
coming from a theoretical model. Kronfeld and Ni\v zi\'c \cite{Kro91}
have for instance calculated  real Compton scattering with various
distribution
 amplitudes as shown on  Figure 12. One sees that the differential Compton
cross section has a high discriminating power with respect to the non
 perturbative object $ \varphi_{(uud)}$ that we want to study.

A less biased way to extract the distribution amplitude from
experimental numbers is to write the cross section as a sum of terms
\beq
A_i T_H^{ij}(\theta) A_j
\eeq
where the decomposition of the distribution amplitude on
 the Appell polynomials (Eq.~(\ref{Appell})) has been used and where
 $T_H^{ij}$ are  integrals over   $x$ and $y$ variables of the product of
the hard amplitude at a given scattering angle
 $\theta$   by the two Appell polynomials
$A_i(x)$ and $A_j(y)$. The $T_H^{ij}$ are ugly long expressions but they
can be numerically handled.

Determining the proton distribution amplitude from experimental data
boils down then to the extraction by a maximum of likelyhood method  of
the  $A_i$ parameters, amputating the series of Eq.~(\ref{Appell}) to its
first $n$  terms, verifying afterwards that including the term $n+1$
does not drastically modify the conclusion. One can then
explore other reactions, virtual Compton scattering for instance,
which must be well described by the same series of $A_i$'s.

\subsection{Other processes }

Photo- and electro-production of mesons at large angle will allow to
probe  distribution amplitudes of $\pi$ and $\rho$ mesons in the same way.
The production of the $ K \Lambda $  final state  selects a few
 hard scattering diagrams. The analysis of these reactions is still to be
done if one excepts some works done in the simplifying framework of the
diquark model~\cite{Kro92}.

Heavy particle decays such as $ B \rightarrow \pi \pi $ have also been
studied in this formalism. We shall not deal here with that interesting physics
case~\cite{CM94}

\section{Independent scattering formalism}

The QCD formalism for the independent scattering process proposed by Landshoff
has been established by  Botts and
Sterman~\cite{BS89}, after pioneering work by Mueller~\cite{Mue81}. A new
factorization property has been derived. An important result is that this
mechanism
asymptotically dominates pure short distance contributions \`a la
Brodsky-Lepage
in hadron-hadron  collisions at fixed angle {\it but is
sub-dominant in the case of  photo- and electro-production  reactions }.

\smallskip
One generally writes an helicity amplitude for the $ \pi \pi \rightarrow \pi
\pi
 $ elastic scattering process of Figure~4 as
\beq
A=\int
\left\{\prod_{i=1}^4 {d^4k_i\over(2\pi)^4} X_{\alpha_i\beta_i}(k_i)\right\}
\left.H(\{k\})H'(\{p-k\})\right|_{\{\alpha\beta\}},
\label{indep}
\eeq
where quark color indices have been skipped,  $\{k\}$ denotes $k_1,k_2,
k_3,k_4$, and only one quark flavor has been kept.

In this equation, $X(k)$ is the Bethe-Salpeter amplitude
\beq
\int d^4ye^{ik.y}
\langle0|T\left(q_{\alpha}(y)~ P(y,0) ~\bar{q}_{\beta}(0)\right)|M(p)\rangle,
\eeq
and $H$ and $H'$ are the subprocesses hard amplitudes, {\it i.e.} a sum of
perturbative QCD graphs.  $H$ is conventionnally the graph where
the quark from meson 1 enters the hard process.

To simplify this expression, we first examine the kinematical regions
which dominate the
integral, either because of the behaviour of various amplitudes, either
because of momentum conservation in the hard diagrams (global momentum
conservation being extracted as usual)
\beq
(2\pi)^4\delta(\textstyle\sum_i k_i)\,(2\pi)^4\delta(\textstyle\sum_i p_i-k_i)
=(2\pi)^4\delta(\textstyle\sum_i k_i)\,(2\pi)^4\delta(\textstyle\sum_i p_i),
\eeq

\subsection{ Kinematics}

It is interesting to attach to each meson $M_i$ a light-cone basis,
 $(v_i,v_i',\xi_i,\eta)$. In the center of  mass system, one chooses
the direction of flight of $M_1$ as the axis $\hat{3}$. Denoting $\theta$
the scattering angle of $M_3$ measured with respect to  $\hat{3}$, one chooses
the axis $\hat{1}$ such that the momentum of $M_3$
be along $\cos\theta\,\hat{3}+\sin\theta\,\hat{1}$. The basis vectors are then
$$\normalbaselineskip 20pt
\matrix{
v_1&\hskip -2mm=\hskip -2mm&
v_2'&\hskip -2mm=\hskip -2mm&
{\displaystyle1\over\displaystyle\sqrt{2}}(\hat{0}+\hat{3}) \hfill
&v_1'&\hskip -2mm=\hskip -2mm&
v_2&\hskip -2mm=\hskip -2mm&
{\displaystyle1\over\displaystyle\sqrt{2}}(\hat{0}-\hat{3}) \hfill\cr
\xi_1&\hskip -2mm=\hskip -2mm&
\xi_2&\hskip -2mm=\hskip -2mm&\hat{1} \hfill
&\eta&\hskip -2mm=\hskip -2mm&\hat{2} \hfill\cr
v_3&\hskip -2mm=\hskip -2mm&
v_4'&\hskip -2mm=\hskip -2mm&
{\displaystyle1\over\displaystyle\sqrt{2}}
        (\hat{0}+\sin\theta\,\hat{1}+\cos\theta\,\hat{3})\ \hfill
&v_3'&\hskip -2mm=\hskip -2mm&v_4&
\hskip -2mm=\hskip -2mm&
{\displaystyle1\over\displaystyle\sqrt{2}}
        (\hat{0}-\sin\theta\,\hat{1}-\cos\theta\,\hat{3}) \hfill\cr
\xi_3&\hskip -2mm=\hskip -2mm&
\xi_4&\hskip -2mm=\hskip -2mm&\cos\theta\,\hat{1}-\sin\theta\,\hat{3};\hfill\cr
}$$
 neglecting the meson  masses in front of
 $Q=\sqrt{s/2}$, the mesons momenta write simply $p_i=Qv_i$.

An analysis similar to the one leading from Eq.~(\ref{virtualite}) to
Eq.~(\ref{virtualiteapprox}) allows one to replace
\beq
H(\{k\})\approx H(\{xQ\})
\eeq
where $x_i$ is the momentum fraction of the quark or the antiquark $i$
 entering the diagram $H$. An equivalent  approximation applies to $H'$.
 One also approximates
\beq
\delta^{(4)}(k_1+k_2-k_3-k_4)\approx{\sqrt{2}\over |\sin\theta|Q^3}
\prod_{i=2}^4\delta(x_1-x_i)\,\delta(l_1+l_2-l_3-l_4),
\eeq
with $l_i$ the momentum carried by the quark or antiquark $i$ along the
direction $\eta$.

This equation shows that all momentum  fractions in $H$ are equal;
 one denotes $x$, the unique resulting fraction, and $\bar{x}=1-x$,
the fraction which prevails in $H'$.

One may then  rearrange  integrals in Eq.~(\ref{indep}),
by introducing the impact parameter $b$
\beq
2\pi\delta(l_i)=\int_{-\infty}^{+\infty}db\,e^{-i(l_3+l_4-l_1-l_2)b},
\eeq
and the hybrid wave function of a meson  propagating along the  $+$ direction,
\beq
{\cal P}_{\alpha\beta}(x,b)=
Q\int\!{dl\over2\pi}e^{ilb}\int\!{dk^-dk^1\over(2\pi)^3}
X_{\alpha\beta}(xQ,k^-,k^1,l),
\eeq
to get
\beq
A(s,t)={\sqrt{2}Q\over2\pi|\sin\theta|}\int_0^1dx\int_{-\infty}^{\infty}db
\Big[H(\{xp\})H'(\{\bar{x}p\})\Big]_{\{\alpha\beta\}}
\prod_{i=1}^4 {{\cal P}_{\alpha_i\beta_i}(x,b;p_i)\over Q}.
\eeq
Each hard process scales like $Q^{-2}$, so that the naive scaling law
for the reaction amplitude is
$\overline{|b|}Q^{-3}$, where $\overline{|b|}$ is a typical average
between the quark and the antiquark in the  valence state of the
mesons. This is the exceptional scale dependence discussed in Section 1.3.

Let us stress that in a short distance convolution, we would have written
\beq
A'=\prod_{i=1}^4\varphi_i(x_i)*T_H(\{x\}),
\eeq
$T_H$ consisting, at the lowest order in the exchange of three hard gluons
for different $x_i$'s. In this convolution,
one gluon becomes soft when all  $x_i$'s become equal and $T_H$
gets an infrared divergence of the type $\int d^4k/k^4$~\cite{Mue81}.

\subsection{ Dynamical factorization and Sudakov suppression }

We already stressed the importance of radiative corrections in
processes involving  hadrons: to evaluate a cross section with a
 perturbative treatment of the theory, one must check that the infrared
regime is under control.

Taking radiative corrections into account modifies the hard process amplitude,
leading to~\cite{BS89}
\beq
\label{amplitude}
A(s,t)={\sqrt{2}Q\over2\pi|\sin\theta|}\int_0^1dx\,H(\{xp\})\,H'(\{\bar{x}p\})
\int_{-1/\Lambda}^{+1/\Lambda}db\,U(x,b,Q)
\prod_{i=1}^4{{\cal P}_i^{(0)}(x,b)\over Q},
\eeq
where the $U$ factor contains the corrections. These corrections turn out
to be very  important by strongly suppressing the integrand in the region
where the impact parameter  $b$  is large in front of the scale $1/Q$.
This is the Sudakov phenomenon~\cite{Sud56} already mentioned above. Then, the
Sudakov-resummed amplitude is still dominated by a short distance dynamics.
Let us here restrict to leading corrections. In axial gauge, they come from
corrections on wave functions. The equation satisfied by $\cal P$
is~\cite{BS89}
\beq
{\partial\over\partial\ln Q}{\cal P}(x,b,Q)=-{1\over2}\left(
\int_{1/b}^{xQ}d\ln\mu'\,\gamma_K+\int_{1/b}^{\bar{x}Q}d\ln\mu'\,\gamma_K
\right){\cal P}(x,b,Q),
\eeq
where
\beq
\gamma_K(\mu')={C_F\over\beta_1\ln\mu'},
\eeq
the solution of which is
\beq
{\cal P}(x,b,Q)={\cal P}^{(0)}(x,b)\,\exp - S(x,b,Q)
\label{P}
\eeq
where
\beq
S(x,b,Q) = \left({c\over4}\ln xQ(u-1-\ln u)
+x\leftrightarrow\bar{x}\right)
\eeq
with $ u(xQ,b)= -{\ln b\over\ln xQ}$,
and  $c=2C_F/\beta_1=32/27$ for three quark flavours.

The generic form of $\cal P$ shows the strong suppression of large transverse
distances $b\gg 1/Q$ (Sudakov suppression) and a regime without corrections
(${\cal P}\approx{\cal P}^{(0)}$) around  $b\sim 1/Q$. In Figure~13
we plot $-S$ as a function of $x$ and $b$.

\vglue5cm
{\centerline \small Figure 13: Exponent of the Sudakov suppression  for the
wave function ($-S(x,b,Q)$) for $Q=2$ and $6\,${\rm GeV}
($\Lambda=200\,${\rm MeV}) as a function of the dipole transverse size $b$
(in {\rm fm}) and of the momentum fraction  $x$.}
\vskip 1cm

One observes that for intermediate values of the energy, the suppression
affects only the region of large transverse distances, but with a very rapid
decrease toward $-\infty$; at higher energies, however,
the correction enforces the process to be dominated by short distances.
Remember that it was a completely different  mechanism, precisely the
physics of
the hard subprocess, which was driving exclusive processes in the short
distance
domain (see section 1).
\smallskip

{\it Back to the pion form factor}

This is where we can come back to the problems noted in section 2.5
concerning the study of the pion form factor at accessible energies .
One may now envisage to compute
the hard scattering without freezing the transverse degrees of freedom
and use the $b$-dependence of the wave function~\cite{Li92}. One gets
(with some technically justified approximations)
\beq
T(-xx'q^2,b)\approx {2\over 3}\pi\alpha_S(t)C_F\,K_0(\sqrt{-xx'q^2}\,b),
\eeq
where $K_0$ is a modified Bessel function.

The interest of this improved approach is that taking radiative corrections
grouped in the wave  function into account, Eq.~(\ref{P}), and analyzing
through the renormalization group the pertinent scale for the coupling
$\alpha_S$ in the above expression of $T$, one gets
\beq
t=\max(1/b,\sqrt{xx'|q^2|}).
\eeq
The Sudakov suppression of large transverse sizes enforces then the form factor
to receive sizeable contributions at large  transfer ($\gapprox 5\,$GeV$^2$),
only from the region where  $b$ is sufficiently small. The scale $t$ of the
perturbative approach then remains large enough in the whole relevant
integration domain.

\medskip
Let us come back to the  Landshoff independent scattering process. The $U$
factor in the amplitude is the product of factors $e^{-S}$ coming from the four
wave functions, {\it i.e.}
\beq
U(x,b,Q)=\exp-\left(c\ln xQ(u-1-\ln u)+x\leftrightarrow\bar{x}\right).
\eeq
We shall not here explicitely calculate the hard  diagrams which are
necessary to  quantitatively compute the amplitude, but simply show how the
$U$ suppression  modifies the counting rule found in
 section 1-3.

Let us thus study the behaviour of the amplitude for $Q\rightarrow\infty$.
At large $Q$ values, one may analytically evaluate the $b$-integral in
Eq.~(\ref{amplitude})
\beq
\int_0^{\Lambda^{-1}}db\,U(b,x,Q),
\eeq
by a saddle point method. To do this, one approximates the exponent in $U$
\beq
c\ln xQ\left(-{\ln b\over\ln xQ}-1-\ln -{\ln b\over\ln xQ}\right)
+x\leftrightarrow\bar{x}
\approx 2c \ln\sqrt{x\bar{x}}Q\,(u-1-\ln u)
\eeq
where $u=-\ln b/\ln\sqrt{x\bar{x}}Q$. Changing then
variables, $b\rightarrow u$, one gets
\beq
\ln\sqrt{x\bar{x}}Q\int_0^{+\infty}du\,
\exp-\ln\sqrt{x\bar{x}}Q\left(2c(u-1-\ln u)+u\right)
\eeq

The exponent is maximum for $u_0={2c \over 2c+1}$ and one gets the approximate
value
\beq
\int_0^{\Lambda^{-1}}db\,U(b,x,Q)\approx
u_0\,\sqrt{\pi \ln Q \over c} (x\bar{x} Q^2)^{c \ln u_0}.
\eeq

The $x$-integration does not modify this behaviour but by  logarithms.
The effect of  radiative corrections is thus to strongly suppress the
contribution
to the  exclusive channel when the impact parameter $b\gg 1/Q$. The
independent interactions must be spatially nearby and the scaling law
is modified as $Q^{-3}\rightarrow Q^{-3.83}$, that is for the differential
cross
section $s^{-5}\rightarrow s^{-5.83}$. In the case of proton-proton elastic
 scattering, the modification is $s^{-8}\rightarrow s^{-9.7}$~\cite{BS89}.

The resulting scaling laws are thus not much different from those deduced for
diagrams respecting the counting rules hypothesis ($s^{-10}$ in the
 $p$-$p$ case).
These two types of  processes are then able to compete and interfere in some
energy interval where their amplitudes are comparable.
This is how one can naturally explain~\cite{PR82} the  experimentally observed
oscillations in the differential cross section as due to the interference
of the amplitudes of these two processes, the Sudakov form factor being
 accompanied by a ``chromo-coulomb"
phase, which  depends logarithmically of the transfer.

\section{ Color transparency}

Hard exclusive scattering ( with a typical large $Q^2$ scale)  selects a very
  special quark configuration: the minimal valence state where
all quarks are  close together,  forming a  small size  color
neutral  configuration  sometimes  referred  to  as a {\em mini hadron}.
This mini hadron  is not a  stationary state and  evolves to build  up a
normal hadron.

Such  a color  singlet system  cannot emit  or absorb  soft gluons
which carry energy or momentum smaller than $Q$.  This is because  gluon
radiation --- like photon radiation in QED --- is a coherent process and
there is thus destructive interference between gluon emission amplitudes
by quarks  with ``opposite''  color.   Even without  knowing exactly how
exchanges  of   soft  gluons   and  other   constituents  create  strong
interactions, we  know that  these interactions  must be  turned off for
small color singlet objects.

An exclusive hard reaction will thus probe the structure of a {\em  mini
hadron}, i.e. the short distance part of a minimal Fock state  component
in the hadron  wave function.   This is of  primordial interest for  the
understanding  of  the  difficult  physics  of  confinement.    First,
selecting the simplest Fock state amounts to the study of the  confining
forces in  a colorless  object which is quite reminiscent of the
``quenched  approximation'' much used in lattice QCD simulations, where
quark-antiquark pair creation from  the vacuum is forbidden.   Secondly,
letting the mini-state evolve during its travel through different nuclei
of various  sizes allows  an indirect  but unique  way to  test how  the
squeezed mini-state  goes back  to its  full size  and complexity,  {\em
i.e.} how  quarks inside  the proton  rearrange themselves  spatially to
``reconstruct'' a normal size hadron.   In this respect the  observation
of baryonic resonance  production as well  as detailed spin  studies are
mandatory.

To the extent that the electromagnetic form factors are understood as  a
function of $Q^2$,
\beq
e+A \rightarrow e+(A-1)+p
\eeq
experiments will measure
the color screening properties of QCD.   The quantity to be measured  is
the transparency ratio $T_r$ which is defined as:
\begin{equation}
T_r = \frac{\sigma_{\rm Nucleus}}{Z \sigma_{\rm Nucleon}}
\end{equation}

At asymptotically large values  of $Q^2$, dimensional estimates  suggest
that $T_r$ scales as a function of $A^{\frac{1}{3}}/Q^2$~\cite{PR91}.
The  approach to the scaling behavior as well as  the value of $T_r$ as a
function  of the  scaling  variable  determine  the  evolution  from  the
pointlike configuration to the  complete hadron. This highly interesting
effect can be measured in an $(e , e' p)$ reaction that provides
the best chance for a {\it quantitative} interpretation. We will not present
here the many ideas which have recently emerged in this new field~\cite{jpr}

\subsection { Present Data}

\hspace {\parindent}
Experimental data on color transparency are very scarce but worth
considering in detail. The first piece of evidence for something like
color transparency came from the Brookhaven experiment on pp elastic
scattering at $90^\circ$  CM in a nuclear medium~\cite{ASC88} .
These data
lead to a lively debate with no unanumous conclusion. The problem is
that hadron hadron elastic scattering is not an as-well clear cut case
of short distance process as the electromagnetic form factors discussed
above. There are indeed infrared sensitive processes (the so-called
independent scattering mechanism) which allow not so small protons to
scatter elastically. The phenomenon of colour transparency is thus
replaced by a {\it nuclear filtering} process: elastic scattering in
a nucleus filters away the big component of the nucleon wave function
and thus its contribution to the cross-section. Since the presence of
these  two competing processes had been analysed \cite{PR82}  as
responsible of the oscillating pattern seen in the scaled cross-section
$s^{10}d\sigma/dt$ , an oscillating color transparency ratio emerges (see
Figure.~14)

\vglue5cm

\centerline{\small Figure 14: Oscillating
 transparency ratio for pp elastic scattering at $90^\circ$
\cite{RP90}.}

One way to understand data is to define a survival probability related in
a standard way to an effective attenuation cross section
 $\sigma_{\rm eff}(Q^2) $ and to plot this attenuation cross section as a
 function of the transfer of the reaction~\cite{JR}. One indeed obtains
 values of  $\sigma_{\rm eff}(Q^2) $ decreasing with $Q^2 $ and quite smaller
 than the free space inelastic proton cross section. The survival probability
is even found to obey a simple scaling law in $Q^2/A^{1/3}$~\cite{PR91}.
The SLAC NE18 experiment~\cite{SLAC} recently measured the color
transparency ratio up to $Q^2=7\,$GeV$^2$ , without any observable increase.
These data are shown in Fig.~15. This casts doubts on the most
optimistic views on very early dominance of pointlike configurations and
emphasizes the importance of a sufficient boost to prevent small states to
dress-up too quickly, then losing their ability to escape freely the nucleus.

\vglue5cm

\centerline{\small Figure 15: The Transparency ratio as measured at
 SLAC\cite{SLAC}}
\vskip1cm

The diffractive electroproduction of vector mesons at Fermilab~\cite{Fermilab}
recently showed an interesting increase of the transparency ratio for data at
$Q^2 \approx 7\,$GeV$^2$. In this case the boost is high since the lepton
energy
is around $E\approx 200\,$GeV, but the problem is to disentangle diffractive
from inelastic events.

\vglue5cm
\centerline{\small Figure 16: The Transparency ratio as measured at
 FNAL~\cite{Fermilab}}
\centerline{\small in diffractive electroproduction of $\rho$}

\subsection{ Future prospects }
It should by now be obvious to the reader that Color Transparency is just an
emerging field of study and that one should devote much attention to get
more information on this physics in the near future.
\subsubsection{ Eva }
A second round of proton experiments at Brookhaven has been approved and
 a new detector named Eva ~\cite{BNL850} with much higher acceptance
 has been taking
data for about one year. Along with other improvements and increased
beam type, this should increase the amount of data taken by a factor
 of 400 allowing a larger energy range and an analysis at different
scattering angles. It would be very interesting to analyze meson-nucleus
 scattering in similar conditions. It has been predicted~\cite{gpr} that
the amount
of helicity non conservation seen for instance in the helicity matrix
elements of the  $\rho$ meson produced in
$\pi p\rightarrow\rho p$ at 90$^\circ$ would be filtered out in a nucleus.
 Experimental data in free space~\cite{hep} yield
$\rho_{1-1}=0.32\pm0.10$, at $s=20.8$GeV$^2$, $\theta_{\rm CM}=90^\circ$,
for the non-diagonal helicity violating matrix element. If the persistence of
helicity non-conservation is correctly understood as due to independent
scattering processes which do not select mini-hadrons and thus are not
subject to color transparency, helicity conservation should be restored
at the same $Q^2$ in processes filtered by  nuclei. One should thus observe
a monotonous decrease of $\rho_{1-1}$ with $A$.

\subsubsection{Hermes}
The Hermes detector ~\cite{HERMES} at HERA is beginning operation. It will
enable a confirmation of FNAL data on $\rho$ meson diffractive production
at moderate $Q^2$ values and quite smaller values of energies
$10 \leq \nu \leq 22\,$GeV. This experiment might however suffer from the
same weakness as the one from Fermilab since Hermes small luminosity only
allows integrated measurements and thus cannot assure that  diffractive
events are not polluted by inelastic events. It seems difficult to envisage in
the near future the detection of the recoiling proton.

\subsubsection{ ELFE }
The $15$--$30\,$GeV continuous electron beam ELFE project has been presented
elsewhere~\cite{ELFE}. Besides the determination of hadronic valence
wave functions through the careful study of many exclusive hard reactions
in free space, the use of nuclear targets to test and use color transparency
is one of its major goals. The $(e,e'p)$ reaction
should in particular be studied in a wide range of $Q^2$ up to $21\,$GeV$^2$,
thus allowing to connect to SLAC data (and better resolution but similar low
$Q^2$ data from CEBAF ) and hopefully clearly establish this phenomenon
in the simplest occurence. The normal component $P_n$ of the polarization
of the recoiling proton will also be
measured in order to discriminate against 2-nucleon knockout and to allow
a limited but fruitful study of shell effects. The vanishing of $P_n$ is
indeed a good signal of the absence of final state interactions.

The measurement of the transparency ratio for photo- and
electroproduction of heavy vector mesons, in particular of $\psi$ and
 $\psi'$ will open  a new regim where the mass of the quark enters as
a new scale controlling the formation length of the produced meson. ELFE
at $30\,$GeV will be the ideal machine to study these physics.

\end{document}